\documentclass[pra,twocolumn,superscriptaddress]{revtex4-1}

\usepackage{braket}
\usepackage{amsmath,bbm}
\usepackage[colorlinks,linkcolor=red,anchorcolor=blue,citecolor=blue]{hyperref}
\usepackage{graphicx}% Include figure files
\usepackage{epstopdf}
\usepackage{dcolumn}% Align table columns on decimal point
\usepackage{bm}% bold math
\usepackage{verbatim}
\usepackage{amsfonts}
\usepackage{color}
\usepackage{multirow}
\usepackage[table,xcdraw]{xcolor}

\usepackage{hyperref}
\usepackage{float}
\usepackage{amssymb,mathrsfs}

\newcommand{\tr}{{\rm Tr}}
\newcommand{\idol}{\ensuremath{\mathbbm 1}}

\begin{document}

\title{Genuine multipartite entanglement verification with convolutional neural networks}
\author{Yi-Jun Luo} %206000337@nbu.edu.cn
\affiliation{School of Physical Science and Technology, Ningbo University, Ningbo 315211, China}
\author{Xuan Leng}
\email{lengxuan@nbu.edu.cn}
\affiliation{School of Physical Science and Technology, Ningbo University, Ningbo 315211, China}
\author{Chengjie Zhang}
\email{chengjie.zhang@gmail.com}
\affiliation{School of Physical Science and Technology, Ningbo University, Ningbo 315211, China}

\begin{abstract}
In recent years, the detection of genuine multipartite entanglement (GME) via machine learning has received scant attention. Here, we employ convolutional neural networks (CNNs), as well as CNNs enhanced with squeeze-and-excitation (SE) to detect GME. We randomly generated GME states with 4 to 6 qubits and GHZ-diagonal states ranging from 4 to 20 qubits using the semidefinite programming approach. Subsequently,  we assessed their classification accuracy. Our results demonstrate that the integration of the SE module significantly improved training performance. Additionally, we conducted an analysis of false positive and false negative occurrences. Utilizing our training data, we have substantially reduced the likelihood of incorrectly classifying non-entangled states as entangled.
%we evaluated their classification accuracy. Our findings indicate that employing the SE module yielded superior training outcomes. Furthermore, we analyzed instances of false positives and false negatives. By leveraging our training results, we effectively mitigate the risk of misclassifying non-entangled states as entangled.
\end{abstract}

%\date{\today}

\maketitle

\section{Introduction}\label{sec:leve1}
In quantum information science, multipartite quantum states are of paramount importance \cite{rev1,rev2,rev3,rev4,rev5,rev6}. These states are fundamental to a variety of communication protocols, such as quantum teleportation \cite{teleportation1,teleportation2,teleportation3}, dense coding \cite{coding1,coding2}, entanglement-based quantum key distribution \cite{key1,key2}, and probing Bell's inequality violations \cite{Bell1,Bell2,Bell3}. Among these, Genuine multipartite entanglement (GME) stands out as the most potent type of entanglement in multipartite quantum systems. Its applications are diverse, ranging from one-way quantum computing using cluster states \cite{cluster}, to exploring quantum phase transitions \cite{phase1,phase2}, and enhancing quantum metrology with Greenberger-Horne-Zeilinger (GHZ) and Dicke states \cite{Dicke1,Dicke2}. Understanding GME is thus crucial for pushing the boundaries of quantum information science.

The task of detecting entanglement is a critical challenge in the realm of quantum entanglement theory \cite{detect1,detect2,detect3,detect4,detect5,detect6}. Recently, machine learning techniques, including neural networks  \cite{network1,network2,network3,network4}, support vector machines \cite{svm}, and safe semi-supervised support vector machines \cite{s4vm}, have emerged as powerful tools for tackling issues in quantum entanglement  \cite{entanglement1,entanglement2,entanglement3,entanglement4,entanglement5} and for verifying quantum steering \cite{steer1,steer2,steer3,steer4}. Despite this progress, research on the machine learning-based detection of GME states remains scarce. Given the crucial role of GME in quantum information processes, which surpasses that of bipartite entanglement, the application of machine learning for its detection is not only promising but also essential.

Currently, a definitive and precise entanglement criterion for GME states remains elusive, often necessitating numerical computations to detect them. As the number of quantum states increases, the form of entanglement becomes more and more complex, posing huge challenges to detection even for special quantum states. In quantum entanglement detection algorithms such as semi-definite programming (SDP) \cite{sdp}, the computational time grows exponentially with dimensionality. In contrast, leveraging machine learning for quantum state detection exhibits polynomial growth in dimensionality, thereby offering substantial savings in time and computational resources. Under the premise of ensuring high accuracy, it has obvious advantages in detecting and generating high-qubit quantum states, and can also provide convenience for quantum computing cloud platforms. Therefore, the endeavor to detect GME through machine learning emerges as both feasible and imperative.

Since the mid-20th century, digital computing has automated data analysis, leading to the emergence and evolution of machine learning. This field has progressed from basic regression \cite{machine1} and principal component analysis \cite{machine2} to sophisticated algorithms like support vector machines. The era of deep learning began in the 1960s, advancing with neural networks \cite{machine3,machine4,machine5} and backpropagation techniques \cite{machine6}. Today's powerful computers, running deep networks with extensive parameters  \cite{machine7,machine8,machine9}, adeptly handle large datasets, revealing complex patterns. Machine learning, a key player in artificial intelligence, enables systems to learn and improve autonomously through statistical and computational principles. Its current applications span quantum steering, nonlocality \cite{nonlocality1,nonlocality2}, and spin systems \cite{spin1,spin2,spin3}, demonstrating its significant promise in quantum information science.

In this study, we explore the application of machine learning techniques for the detection of GME states. We commence by introducing convolutional neural networks (CNNs) \cite{cnn} with an integrated squeeze-and-excitation (SE) module \cite{se}, designed to enhance performance. We then assess the algorithm's predictive accuracy in identifying randomly generated GME states across 4 to 6 qubits and GHZ-diagonal states for 4 to 20 qubits, both pre- and post-enhancement. The correct GME states are labeled using SDP, and Corollary 7 from the referenced literature aids in \cite{ghz_d} identifying the accurate GHZ-diagonal states. Our results reveal that the augmented CNN algorithm significantly surpasses the conventional model, achieving an average classification accuracy of over 92$\%$ for GME states and in excess of 98$\%$ for GHZ-diagonal states. Notably, as the qubit count rises, the algorithm consistently maintains high accuracy. Additionally, we examine instances of false positives and negatives, concluding that our model exhibits superior accuracy in predicting entangled states compared to non-entangled states.

% The paper is organized as follows. Section II presents the SDP algorithm, CNN algorithm and SE module. In Sec. III, we study the use of CNN algorithms before and after enhancement to detect GME states and GHZ-diagonal states, and analyze the results. Last but not least, Section IV concludes the study.

\section{\label{sec:leve2}Methods}
\subsection{\label{sec:leve21}Detecting GME using the SDP Method}
The SDP algorithm serves as a technique for resolving quadratic optimization problems by transforming the original problem into an SDP problem through positive semidefinite relaxation.

For a bipartition $\alpha|\bar{\alpha}$, we employ an entanglement witness $W=P_{\alpha}+Q_{\alpha}^{T_{\alpha}}$, comprising positive operators $P_{\alpha}$ and $Q_{\alpha}$, to define an entanglement monotone for quantifying GME, namely the genuine multipartite negativity (GMN) \cite{gmn}. The GMN can be computed via SDP, and a renormalized version of GMN has been proposed \cite{ghz_d}. For a qubit state $\rho$, the renormalized GMN $N_g(\rho)$ is expressed as
\begin{equation}
	\begin{aligned}
		& \ \ \ N_g(\rho)=  -  \inf \tr(W \rho) \\
		& \text { subject to: }   W=P_{\alpha}+Q_{\alpha}^{T_{\alpha}} \\
		& \ \ \ \ \ \  \ \ \ \ \ \ \ \ \ \ \ \  0\leq P_{\alpha} \\
		& \ \ \ \ \ \  \ \ \ \ \ \ \ \ \ \ \ \  0\leq Q_{\alpha}\leq \idol \text{ for all bipartitions } \alpha|\bar{\alpha},
	\end{aligned}\label{eq.1}
\end{equation}
where $\alpha$ traverses all possible subsystems, and $T_{\alpha}$ denotes the partial transpose for subsystem $\alpha$. Notably, the renormalized GMN equals a mixed-state convex roof of bipartite negativity,
\begin{eqnarray}\label{GMN}
	N_g(\varrho)=\inf_{p_\alpha,\varrho_{\alpha}}\sum_\alpha p_\alpha N_{\alpha}(\varrho_{\alpha}),
\end{eqnarray}
where the summation covers all possible decompositions $\alpha|\overline{\alpha}$ of the system, and the minimization occurs over all mixed state decompositions of the state $\varrho=\sum_\alpha p_\alpha \varrho_{\alpha}$.

Utilizing SDP (\ref{eq.1}), we can randomly generate n-qubit quantum states and derive their renormalized GMN values. If the renormalized GMN value is positive, we consider the quantum state to be genuinely entangled and label it as $-1$. If the target value is non-positive, we label it as $+1$. To ensure balanced data, we generate an equal number of samples from both classes.

\subsection{\label{sec:leve22}Convolutional Neural Networks}
CNN is a deep feed-forward neural network characterized by local connections and weight sharing, representing a cornerstone algorithm in deep learning. One of its key advantages lies in its ability to reduce the dimensionality of a large parameter set into a smaller one (through convolution operations), thereby simplifying complex problems. Widely employed in tasks such as image classification, target detection, and image segmentation, CNN stands as one of the most extensively used models. Therefore, leveraging CNN for the detection of entangled states allows effective extraction of both local and global characteristics, demonstrating robust applicability in the detection of high-dimensional entangled quantum states.

Once CNN receives data from the input layer, it undergoes processing through convolution, pooling, and other layers to extract and transform features. Subsequently, the fully connected layer aggregates these features to produce the final output.

\textit{Input Layer}: Typically, CNN receives original or preprocessed data as input. In the context of image processing, where input data usually comprises colorful images with three RGB channels, the input matrix size is $H \times W \times 3$. However, when dealing with quantum states, we obtain a density matrix with only one channel, resulting in an input matrix size of $H \times H \times 1$, where $H$ corresponds to the dimension of the Hilbert space. Through a separation method, we divide the density matrix into real and imaginary parts, subsequently eliminating zero values on the diagonal of the imaginary matrix to yield an input matrix size of $(2 \times H-1) \times H \times 1$. For GHZ-diagonal states, we acquire an input matrix size of $H \times 1 \times 1$, as discussed later.

\begin{figure*}[t]
	\centering
	% Requires \usepackage{graphicx}
	\includegraphics[width=7in]{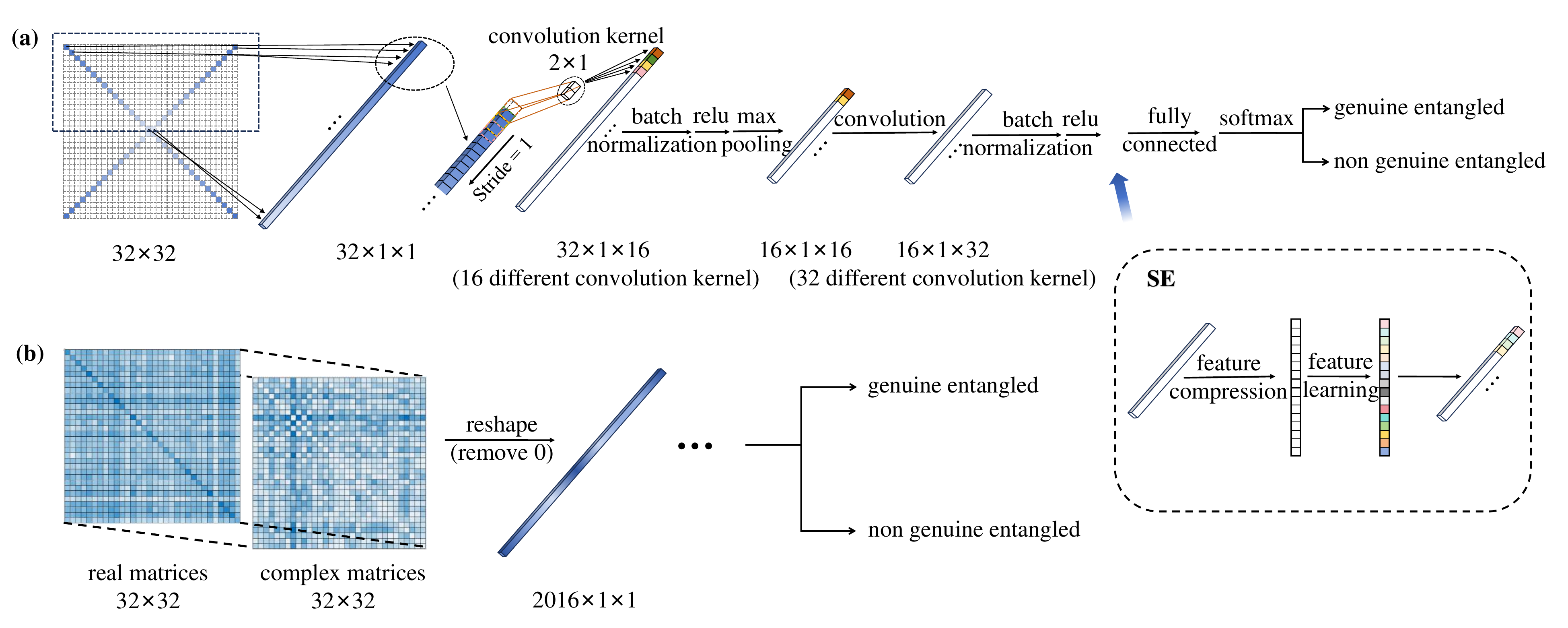}
	\caption{Diagram depicting the CNN network structure and the preprocessing steps for GHZ-diagonal states (a) and GME states (b), where the color variations do not signify specific values but are solely for comparison of data size for better comprehension.}\label{FIG.1}
\end{figure*}

\textit{Convolutional Layer}: The pivotal operation in CNN is convolution. Through this operation, matrix features are extracted, and the matrix size is simplified, thereby enhancing computational efficiency. Typically, since the original matrix is two-dimensional, CNN employs 2D convolution, allowing the kernel to move in the $x$ and $y$ directions without crossing channels. However, we preprocess the density matrix into a 1D form, facilitating the use of 1D convolution, which moves solely along the $x$ axis, offering greater convenience than 2D convolution. This is because CNN is often used for image classification. Suppose there is a $32 \times 32$ image, in which an item contains information of $4 \times 4$ size, then it is more effective and accurate to use 2D convolution to obtain its image features. However, the distribution of entangled information in the density matrix is not exactly the same as the distribution of image features in image recognition. For density matrices, the information usually containing entanglement is relatively scattered, and the global impact needs to be considered. It is impossible to detect whether it is entangled through a small part of the information, so the effect of using 2D convolution is not necessarily better than 1D convolution. If there is a GHZ-diagonal state, after reshaping the density matrix, only the values on the diagonal are retained. Using a $2 \times 1$ convolution kernel for 1D convolution can largely retain the diagonal features. If 2D convolution is used, the convolution kernel is generally larger than $2 \times 1$, so more information is lost than one-dimensional convolution, which will affect the accuracy results. In addition, if 2D convolution is used when facing high-dimensional matrices, it may face the problem of insufficient memory and it may be difficult to retain the diagonal characteristics of the density matrix.

In principle, the convolution operation involves a mathematical process of dot multiplication and summation between two pixel matrices. One matrix represents the input data, while the other matrix is the convolution kernel, also known as the filter or feature matrix. The resulting output signifies the local features extracted through feature extraction from the original image. In practical implementation, as depicted in Fig. \ref{FIG.1}, we employ zero padding at the top and valid convolution at the bottom. This is because when applying a $2\times1$ convolution kernel to an image of size $32\times1$, the resulting feature map dimension is $31\times1$. To maintain consistency with the original image size, it is common practice to prepend a zero before the first term, effectively expanding the input image size to $33\times1$. This approach preserves spatial alignment and facilitates proper convolutional operations across the entire image. Furthermore, the convolutional layer may comprise multiple kernels. In real-world scenarios, to obtain $n$ features, multiple convolution kernels are often utilized. Here, we employ 16 and 32 distinct convolution kernels, resulting in corresponding matrix sizes of $(2 \times H-1) \times H \times 16$ and $(2 \times H-1)/2 \times H \times 32$, respectively.

\textit{Batch Normalization Layer}: The batch normalization layer employs standardization processing to address numerical instability issues in CNNs, ensuring a more uniform feature distribution in this layer, which aids subsequent network training.

\textit{Activation Layer}: The activation function is pivotal in the activation layer, providing the network with nonlinear modeling capabilities. As discussed in the introduction to the convolutional layer, the convolution method processes data, constituting a linear operation. However, not all samples are linearly separable. To address this, we introduce nonlinear operations, i.e., activation functions, to enable the model to handle nonlinear classification problems. Without an activation function, the network can only express linear mappings, rendering it equivalent to a single-layer neural network and essentially a linear model, even with multiple hidden layers. Hence, through the activation function, the CNN convolutional neural network acquires nonlinear mapping capabilities. Common activation functions include the sigmoid function: $\sigma(x)=\frac{1}{1+e^{-x}}$, the tanh function: $\tanh (x)=\frac{\sinh (x)}{\cosh (x)}=\frac{e^x-e^{-x}}{e^x+e^{-x}}$, and the ReLU function: $\operatorname{ReLU}(x)=\max (x, 0)$, etc. Here, we utilize the ReLU function. Its advantage lies in maintaining a gradient of 1 when $x>0$, thereby avoiding the vanishing gradient problem and ensuring rapid convergence.

By employing the ReLU function, we enhance the sparsity of the network, facilitating the prominence of representative features and strengthening the model's generalization ability; fewer neurons can achieve the same effect in training. However, excessive sparsity induced by the ReLU function may impede effective feature learning. It's worth noting that in practical applications, the convolutional layer, batch normalization layer, and ReLU layer can be interconnected. As depicted in Fig. \ref{FIG.1}, we integrate two such modules, which enhance the model's training capabilities.

\textit{Pooling Layer}: Pooling constitutes a downsampling operation aimed at reducing the feature space size while retaining essential information. This technique not only reduces computational complexity but also enhances the model's generalization ability and mitigates the risk of overfitting.

Pooling layers come in three main types: max pooling, average pooling, and global pooling. In order to compare which pooling is better for CNN, we tried using average pooling instead of max pooling on the CNN and got similar results. However, after the attention mechanism is introduced, the results of max pooling are a little better than those obtained by average pooling, so we adopt max pooling for CNN both before and after introducing attention. Max pooling is the most prevalent method, ,and we employ it here. Through max pooling, the most significant features are preserved, as local maxima often represent salient features. In contrast, average pooling computes the average of local area features to maintain overall features. The global pooling layer, typically used at the end of the network, reduces each feature map to a single value, significantly reducing model parameters and preventing overfitting.

\textit{Fully Connected Layer}: Serving as the ``classifier" in CNN, the fully connected layer processes previously extracted feature maps. It multiplies the input feature vector by the weight matrix obtained during training, adds the bias term, and applies nonlinear mapping to yield the final classification result. Here, we utilize the Softmax function:
\begin{equation}
	P(y=j)=\frac{e^{x^T} W_j}{\sum_{k=1}^K e^{x^T W_k}},
\end{equation}
where $K$ denotes the different types of linear functions, $x$ represents the sample vector, and $j$ signifies the classification category.

\textit{Adaptive Moment estimation method (Adam)}: Adam \cite{adam1} computes adaptive learning rates for each parameter. Combining the Momentum gradient descent method \cite{adam2} and the RMSprop algorithm \cite{adam3}, Adam stands as one of the few optimization algorithms suitable for various deep learning tasks. It adjusts each parameter's learning rate by computing first-order and second-order moment estimates of the gradient. By incorporating momentum terms to accelerate training and constraining learning rates, Adam ensures efficient network training. It is defined as:
\begin{equation}
	\left\{
	\begin{aligned}
		&\hat{\mathrm{m}}_{\mathrm{t}}=\frac{\beta_1 \mathrm{~m}_{\mathrm{t}-1}+\left(1-\beta_1\right) \mathrm{g}_{\mathrm{t}}}{1-\beta_1^{\mathrm{t}}}\\
		&\hat{\mathrm{v}}_{\mathrm{t}}=\frac{\beta_2 \mathrm{v}_{t-1}+\left(1-\beta_2\right)\mathrm{g}_t^2}{1-\beta_2^t}\\
		&\theta_{\mathrm{t}+1}=\theta_{\mathrm{t}}-\frac{\eta}{\sqrt{\hat{\mathrm{v}}_{\mathrm{t}}}+\epsilon} \hat{\mathrm{m}}_{\mathrm{t}}
	\end{aligned}
	\right.
\end{equation}
where $\hat{\mathrm{m}}_{\mathrm{t}}$ and $\hat{\mathrm{v}}_{\mathrm{t}}$ are bias-corrected first and second-order moment estimates, $\beta_1$ and $\beta_2$ are the exponential decay coefficients of the two moment estimates, usually set to 0.9 and 0.999, $\eta$ is the learning rate, which is usually not fixed and can be set to 0.001 in most cases, $\epsilon$ is a very small number, the value is 1e-8, and $\theta_{\mathrm{t}+1}$ denotes the updated parameter. Adam outperforms other adaptive learning rate algorithms, offering faster convergence, more effective learning, and addressing issues like vanishing learning rates, slow convergence, or high variance parameter updates leading to large fluctuations in the loss function.

\subsection{\label{sec:leve23}Squeeze and Excitation}
Squeeze and Excitation (SE) constitutes a channel attention module designed to enhance channel features within the input feature map without altering its size. The fundamental concept involves network-driven learning to adaptively allocate weights to different channels. This characteristic facilitates the enhancement of feature learning, thereby improving the network's capacity to extract entangled information from the density matrix. Moreover, the SE attention mechanism exhibits high adaptability, allowing optimization tailored to diverse tasks and datasets. Its autonomous training via networks minimizes manual intervention, thereby augmenting both model performance and generalization capabilities.

\begin{figure}[t]
	\centering
	% Requires \usepackage{graphicx}
	\includegraphics[width=3.5in]{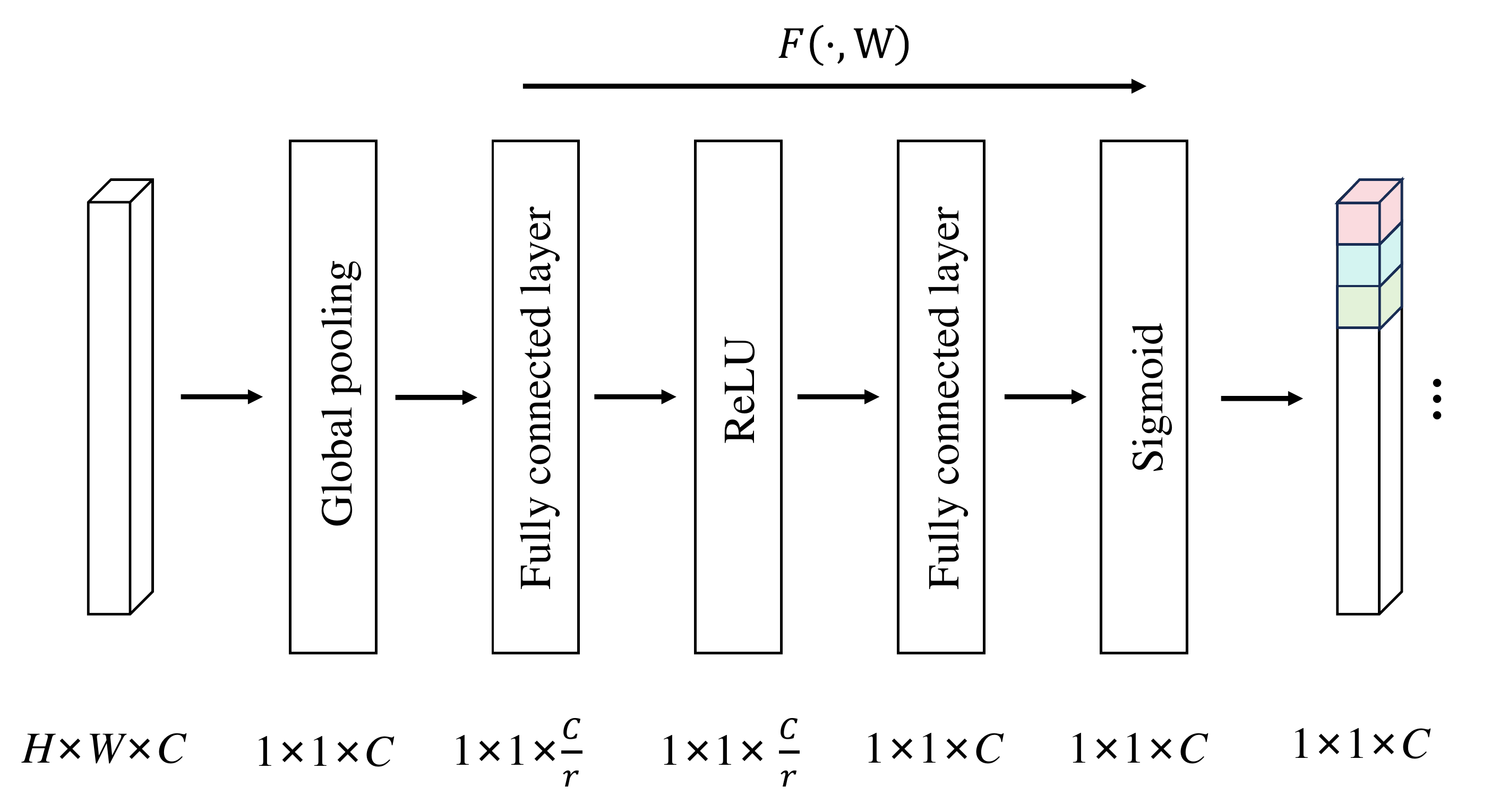}\\
	\caption{The structural diagram of SE.}\label{FIG.2}
\end{figure}

To leverage channel dependencies, the first step involves global average pooling across channels, compressing global information into a single channel. Consequently, the $H \times W \times C$ feature map transforms into the $1 \times 1 \times C$ feature vector $z$, as depicted in Fig. \ref{FIG.2}. This compresses the channel features of $C$ feature maps into a single numerical value per channel in the feature vector $z$, formulated as:
\begin{equation}
	z_c=\mathbf{F}_{sq}(\mathbf{u}_c)=\frac{1}{H \times W} \sum_{i=1}^H \sum_{j=1}^W u_c(i, j)
\end{equation}
where $z_c$ represents the $c$-th element of $Z$.

\textit{Excitation}: This phase entails a specialized module comprising two fully connected layers, aiming to apply two nonlinear functions: ReLU and sigmoid. The first fully connected layer compresses $C$ channels into $\frac{C}{r}$ channels to reduce computational load, followed by a ReLU nonlinearity, where $r$ denotes the compression ratio. Subsequently, the second fully connected layer restores the channel count to $C$ channels, subsequently passing through a sigmoid function. This yields a weight vector $s$ of dimension $1 \times 1 \times C$, expressed as:
\begin{equation}
	\mathbf{s}=\mathbf{F}_{ex}(\mathbf{z}, \mathbf{W})=\sigma(g(\mathbf{z}, \mathbf{W}))=\sigma\left(\mathbf{W}_2 \delta\left(\mathbf{W}_1 \mathbf{z}\right)\right)
\end{equation}

Finally, the obtained attention weights are applied to the features of each channel, yielding:
\begin{equation}
	\widetilde{\mathbf{x}}_c=\mathbf{F}(\mathbf{u}_c, s_c)=s_c \mathbf{u}_c.
\end{equation}

%\begin{table}[t]
%	%\centering
%	\renewcommand\arraystretch{1.5}
%	\tabcolsep=0.205cm
%	\begin{tabular}{l*{4}c}
%\hline
%\hline
%		%\cline{1-4}
%		$l$  & S4VM & SVM-S4VM & Renewal SVM-S4VM &  \\ \hline% \cline{1-4}
%		40 & 78.76 & 82.17    & 86.16       &  \\
%		60 & 82.30  & 85.47    & 89.72       &  \\
%		80 & 82.80  & 85.91    & 90.52       &  \\ %\cline{1-4}
%\hline
%\hline
%	\end{tabular}
%	\caption{The average maximum prediction accuracy ($\%$) of the S4VM, SVM-S4VM and Renewal SVM-S4VM methods with $l=40$, $60$, and $80$.}\label{tab.2}
%\end{table}

\section{NUMERICAL RESULTS}\label{sec:leve3}
\subsection{\label{sec:leve31}Genuine multipartite entanglement}
A state $\varrho^{\text {bs}}$ that is not fully separable is called biseparable if it can be written as a convex combination of biseparable pure states:
\begin{equation}\label{eq.8}
	\varrho^{\mathrm{bs}}=\sum_i p_i\left|\phi_i^{\mathrm{bs}}\right\rangle\left\langle\phi_i^{\mathrm{bs}}\right|.
\end{equation}
Note that the biseparable states $\left|\phi_i^{\text {bs }}\right\rangle$ might be biseparable with respect to different partitions. If the quantum state cannot be expressed in Eq.~(\ref{eq.8}), it is a GME state \cite{rev3}.

We employed the SDP program, implemented through YALMIP with MOSEK, to randomly generate 4, 5, and 6-qubit quantum states, obtained their renormalized GMN values, and determined their entanglement status. To ensure balance in the samples, we generated 1500 quantum states with $-1$ labels and 1500 quantum states with $+1$ labels for each of these three qubit types. Attempts were made to generate a 7-qubit quantum state, but obtaining its corresponding GMN value through the SDP method proved challenging due to the considerable memory and time resources required.

Prior to CNN training, we randomly shuffled the entangled state and non-entangled state samples. Among these, 1050 $-1$ labeled samples and 1050 $+1$ labeled samples were allocated for the training set, accounting for 70$\%$ of the total data, while the remaining samples were used for the test set. In CNN, alongside the parameter settings mentioned earlier, we set MaxEpochs to 200, InitialLearnRate to 0.001, applied L2Regularization of 1e-4, and shuffled the dataset for each training iteration.

\begin{table}[t]
	\centering
	\renewcommand\arraystretch{1.5}
	\tabcolsep=0.65cm
	\caption{Classification accuracy ($\%$) of CNN experimental results with and without SE module.}\label{tab.1}
	
	\begin{tabular}{l*{4}c}
		\hline
		\hline
		qubit     & 4        & 5  & 6  \\\hline
		CNN      & 96.67      & 92.33 & 89.33 \\
		CNN-SE      & 98.33      & 94.33 & 92.00 \\
		$\Delta_+$    & 1.66    & 2.00  & 2.67 \\
		\hline
		\hline
	\end{tabular}
\end{table}

In Table \ref{tab.1}, we compare the training results of CNN and CNN with the SE module (CNN-SE) across 5 repeated experiments. It was observed that as the number of qubits increases, the classification accuracy slowly decreases, reaching 98.33$\%$ at 4-qubit and 92$\%$ accuracy at 6-qubit. We speculate that this decrease in accuracy is related to the increase in training data. The density matrix of a 6-qubit system is much larger than that of a 4-qubit system, making it more challenging for machine learning algorithms to effectively learn classification features.

Furthermore, we found that the classification accuracy of CNN-SE is consistently higher than that of CNN, with an improvement of approximately 2$\%$. This improvement increases with the number of qubits. We attribute this to the fact that CNN achieves higher training accuracy at lower qubit counts, leaving less room for improvement. Based on these observations, we speculate that beyond 6-qubit systems, our CNN-SE model will likely outperform the CNN model and exhibit greater potential.

\begin{figure}[t]
	\centering
	% Requires \usepackage{graphicx}
	\includegraphics[width=3.5in]{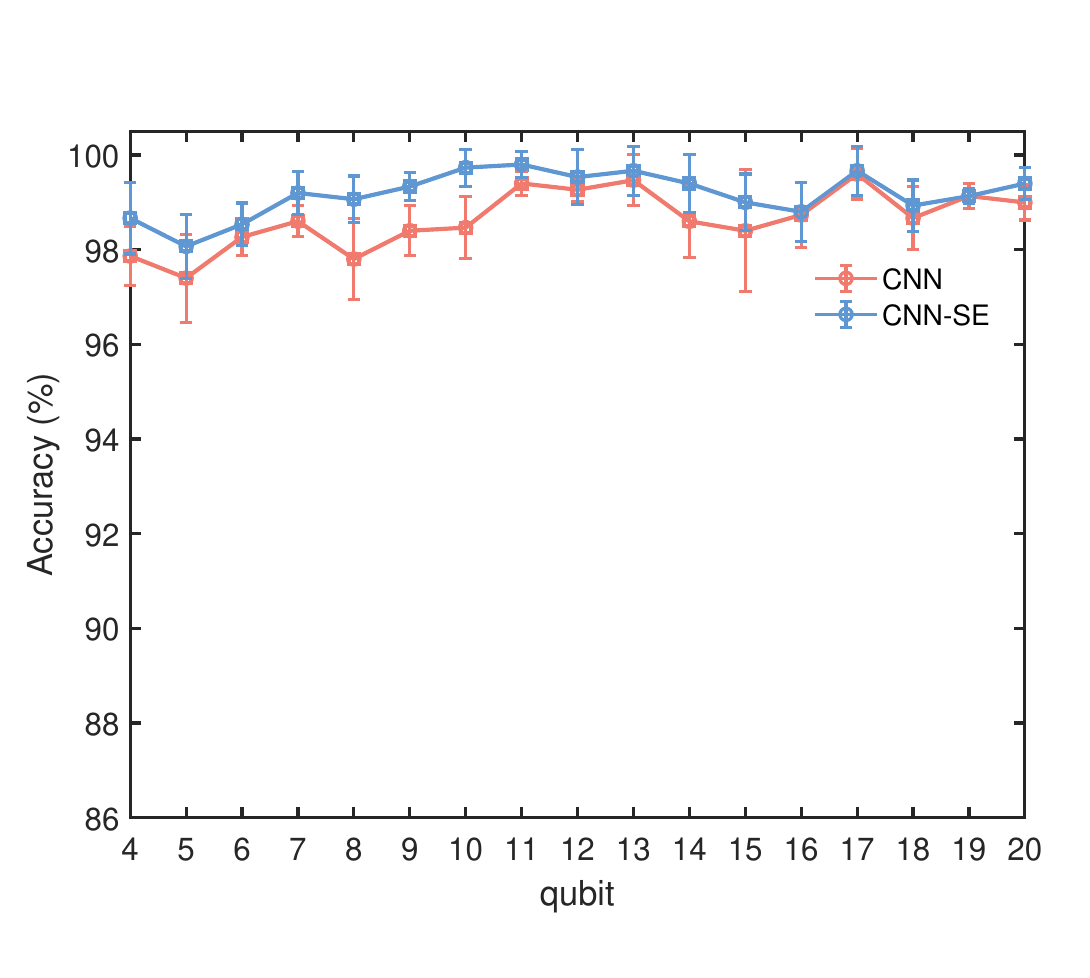}\\
	\caption{Classification accuracy and error bars for different numbers of qubits obtained through CNN and CNN-SE.}\label{FIG.3}
\end{figure}

\subsection{\label{sec:leve32}n-qubit GHZ-diagonal states}
\begin{figure*}[t]
	\centering
	% Requires \usepackage{graphicx}
	\includegraphics[width=7in]{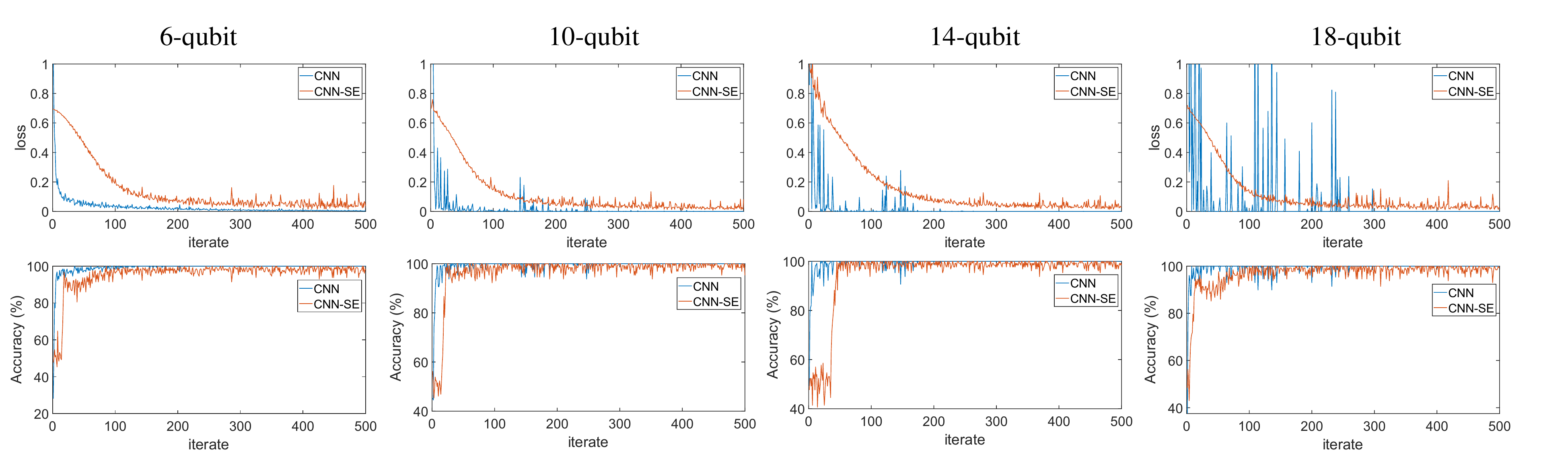}\\
	\caption{From left to right are the corresponding loss functions and accuracy training curves for 6-qubit, 10-qubit, 14-qubit, and 18-qubit during the training process of CNN and CNN-SE.}\label{FIG.4}
\end{figure*}

\begin{table*}[t]
	\caption{Number of classification errors in CNN experimental results with and without SE module.}\label{tab.2}
	\resizebox{1\textwidth}{!}{%
		\centering
		\begin{tabular}{|c|l|lllll|lllll|lllll|lllll|lllll|}
			\hline
			\multicolumn{1}{|l|}{}            &                                 & \multicolumn{5}{c|}{\cellcolor[HTML]{DAE8FC}9-qubit}                                                  & \multicolumn{5}{c|}{\cellcolor[HTML]{DAE8FC}10-qubit}                                                 & \multicolumn{5}{c|}{\cellcolor[HTML]{DAE8FC}11-qubit}                                                 & \multicolumn{5}{c|}{\cellcolor[HTML]{DAE8FC}12-qubit}                                                 &
			\multicolumn{5}{c|}{\cellcolor[HTML]{DAE8FC}13-qubit}												\\ \hline
			& entangled predict non-entangled & \multicolumn{1}{l|}{7} & \multicolumn{1}{l|}{3} & \multicolumn{1}{l|}{3} & \multicolumn{1}{l|}{0} & 2 & \multicolumn{1}{l|}{0} & \multicolumn{1}{l|}{1} & \multicolumn{1}{l|}{3} & \multicolumn{1}{l|}{3} & 5 & \multicolumn{1}{l|}{2} & \multicolumn{1}{l|}{1} & \multicolumn{1}{l|}{0} & \multicolumn{1}{l|}{1} & 1 & \multicolumn{1}{l|}{0} & \multicolumn{1}{l|}{3} & \multicolumn{1}{l|}{2} & \multicolumn{1}{l|}{1} & 2 & \multicolumn{1}{l|}{0} & \multicolumn{1}{l|}{3} & \multicolumn{1}{l|}{1} & \multicolumn{1}{l|}{0} & 2\\ \cline{2-27}
			\multirow{-2}{*}{\textbf{CNN}}    & non-entangled predict entangled & \multicolumn{1}{c|}{3} & \multicolumn{1}{l|}{3} & \multicolumn{1}{l|}{2} & \multicolumn{1}{l|}{3} & 1 & \multicolumn{1}{l|}{2} & \multicolumn{1}{l|}{3} & \multicolumn{1}{l|}{1} & \multicolumn{1}{l|}{2} & 3 & \multicolumn{1}{l|}{0} & \multicolumn{1}{l|}{3} & \multicolumn{1}{l|}{3} & \multicolumn{1}{l|}{0} & 0 & \multicolumn{1}{l|}{2} & \multicolumn{1}{l|}{0} & \multicolumn{1}{l|}{1} & \multicolumn{1}{l|}{0} & 0 & \multicolumn{1}{l|}{0} & \multicolumn{1}{l|}{1} & \multicolumn{1}{l|}{0} & \multicolumn{1}{l|}{0} & 1\\ \hline
			& entangled predict non-entangled & \multicolumn{1}{l|}{1} & \multicolumn{1}{l|}{0} & \multicolumn{1}{l|}{2} & \multicolumn{1}{l|}{0} & 1 & \multicolumn{1}{l|}{0} & \multicolumn{1}{l|}{0} & \multicolumn{1}{l|}{0} & \multicolumn{1}{l|}{1} & 1 & \multicolumn{1}{l|}{1} & \multicolumn{1}{l|}{0} & \multicolumn{1}{l|}{2} & \multicolumn{1}{l|}{0} & 0 & \multicolumn{1}{l|}{1} & \multicolumn{1}{l|}{1} & \multicolumn{1}{l|}{0} & \multicolumn{1}{l|}{0} & 0 & \multicolumn{1}{l|}{1} & \multicolumn{1}{l|}{1} & \multicolumn{1}{l|}{0} & \multicolumn{1}{l|}{0} & 0\\ \cline{2-27}
			\multirow{-2}{*}{\textbf{CNN-SE}} & non-entangled predict entangled & \multicolumn{1}{l|}{2} & \multicolumn{1}{l|}{1} & \multicolumn{1}{l|}{0} & \multicolumn{1}{l|}{3} & 0 & \multicolumn{1}{l|}{0} & \multicolumn{1}{l|}{0} & \multicolumn{1}{l|}{0} & \multicolumn{1}{l|}{2} & 0 & \multicolumn{1}{l|}{0} & \multicolumn{1}{l|}{0} & \multicolumn{1}{l|}{0} & \multicolumn{1}{l|}{0} & 0 & \multicolumn{1}{l|}{3} & \multicolumn{1}{l|}{2} & \multicolumn{1}{l|}{0} & \multicolumn{1}{l|}{0} & 0 & \multicolumn{1}{l|}{0} & \multicolumn{1}{l|}{3} & \multicolumn{1}{l|}{0} & \multicolumn{1}{l|}{0} & 0 \\ \hline
		\end{tabular}
	}
	\\
	\resizebox{1\textwidth}{!}{%
		\centering
		\begin{tabular}{|c|lllll|lllll|lllll|lllll|lllll|lllll|lllll|}
			\hline
			\multicolumn{1}{|l|}{}            & \multicolumn{5}{c|}{\cellcolor[HTML]{DAE8FC}\fontsize{8}{10}\selectfont 14-qubit}                                                 & \multicolumn{5}{c|}{\cellcolor[HTML]{DAE8FC}\fontsize{8}{10}\selectfont 15-qubit}                                                 & \multicolumn{5}{c|}{\cellcolor[HTML]{DAE8FC}\fontsize{8}{10}\selectfont 16-qubit}                                                 & \multicolumn{5}{c|}{\cellcolor[HTML]{DAE8FC}\fontsize{8}{10}\selectfont 17-qubit}                                                 &
			\multicolumn{5}{c|}{\cellcolor[HTML]{DAE8FC}\fontsize{8}{10}\selectfont 18-qubit}												  &
			\multicolumn{5}{c|}{\cellcolor[HTML]{DAE8FC}\fontsize{8}{10}\selectfont 19-qubit}												  &
			\multicolumn{5}{c|}{\cellcolor[HTML]{DAE8FC}\fontsize{8}{10}\selectfont 20-qubit}											\\ \hline
			& \multicolumn{1}{l|}{4} & \multicolumn{1}{l|}{0} & \multicolumn{1}{l|}{6} & \multicolumn{1}{l|}{0} & 4 & \multicolumn{1}{c|}{0}  & \multicolumn{1}{l|}{5} & \multicolumn{1}{l|}{2} & \multicolumn{1}{l|}{4} & 0 & \multicolumn{1}{l|}{1} & \multicolumn{1}{l|}{0} & \multicolumn{1}{l|}{5} & \multicolumn{1}{l|}{0} & 3 & \multicolumn{1}{l|}{0} & \multicolumn{1}{l|}{0} & \multicolumn{1}{l|}{0} & \multicolumn{1}{l|}{0} & 2 & \multicolumn{1}{l|}{7} & \multicolumn{1}{l|}{5} & \multicolumn{1}{l|}{3} & \multicolumn{1}{l|}{1} & 2 & \multicolumn{1}{l|}{3} & \multicolumn{1}{l|}{3} & \multicolumn{1}{l|}{3} & \multicolumn{1}{l|}{0} & 1 & \multicolumn{1}{l|}{4} & \multicolumn{1}{l|}{3} & \multicolumn{1}{l|}{2} & \multicolumn{1}{l|}{1} & 0 \\ \cline{2-36}
			\multirow{-2}{*}{\textbf{{\fontsize{7.6}{10}\selectfont CNN}}}    & \multicolumn{1}{c|}{0} & \multicolumn{1}{l|}{3} & \multicolumn{1}{l|}{1} & \multicolumn{1}{l|}{0} & 1 & \multicolumn{1}{l|}{12} & \multicolumn{1}{l|}{0} & \multicolumn{1}{l|}{0} & \multicolumn{1}{l|}{0} & 1 & \multicolumn{1}{l|}{0} & \multicolumn{1}{l|}{3} & \multicolumn{1}{l|}{0} & \multicolumn{1}{l|}{7} & 0 & \multicolumn{1}{l|}{0} & \multicolumn{1}{l|}{0} & \multicolumn{1}{l|}{0} & \multicolumn{1}{l|}{4} & 0 & \multicolumn{1}{l|}{0} & \multicolumn{1}{l|}{0} & \multicolumn{1}{l|}{0} & \multicolumn{1}{l|}{0} & 2 & \multicolumn{1}{l|}{0} & \multicolumn{1}{l|}{0} & \multicolumn{1}{l|}{0} & \multicolumn{1}{l|}{1} & 2 & \multicolumn{1}{l|}{1} & \multicolumn{1}{l|}{0} & \multicolumn{1}{l|}{1} & \multicolumn{1}{l|}{1} & 2 \\ \hline
			& \multicolumn{1}{l|}{0} & \multicolumn{1}{l|}{2} & \multicolumn{1}{l|}{1} & \multicolumn{1}{l|}{2} & 0 & \multicolumn{1}{c|}{4}  & \multicolumn{1}{l|}{5} & \multicolumn{1}{l|}{4} & \multicolumn{1}{l|}{2} & 0 & \multicolumn{1}{l|}{1} & \multicolumn{1}{l|}{5} & \multicolumn{1}{l|}{2} & \multicolumn{1}{l|}{6} & 0 & \multicolumn{1}{l|}{3} & \multicolumn{1}{l|}{0} & \multicolumn{1}{l|}{0} & \multicolumn{1}{l|}{0} & 1 & \multicolumn{1}{l|}{0} & \multicolumn{1}{l|}{2} & \multicolumn{1}{l|}{0} & \multicolumn{1}{l|}{3} & 0 & \multicolumn{1}{l|}{2} & \multicolumn{1}{l|}{3} & \multicolumn{1}{l|}{0} & \multicolumn{1}{l|}{0} & 1 & \multicolumn{1}{l|}{2} & \multicolumn{1}{l|}{1} & \multicolumn{1}{l|}{0} & \multicolumn{1}{l|}{0} & 3 \\ \cline{2-36}
			\multirow{-2}{*}{\textbf{{\fontsize{7.6}{10}\selectfont CNN-SE}}} & \multicolumn{1}{l|}{0} & \multicolumn{1}{l|}{3} & \multicolumn{1}{l|}{1} & \multicolumn{1}{l|}{0} & 0 & \multicolumn{1}{c|}{0}  & \multicolumn{1}{c|}{0} & \multicolumn{1}{l|}{0} & \multicolumn{1}{l|}{0} & 0 & \multicolumn{1}{l|}{3} & \multicolumn{1}{l|}{0} & \multicolumn{1}{l|}{0} & \multicolumn{1}{l|}{0} & 1 & \multicolumn{1}{l|}{1} & \multicolumn{1}{l|}{0} & \multicolumn{1}{l|}{0} & \multicolumn{1}{l|}{0} & 0 & \multicolumn{1}{l|}{5} & \multicolumn{1}{l|}{0} & \multicolumn{1}{l|}{5} & \multicolumn{1}{l|}{0} & 1 & \multicolumn{1}{l|}{1} & \multicolumn{1}{l|}{0} & \multicolumn{1}{l|}{2} & \multicolumn{1}{l|}{3} & 1 & \multicolumn{1}{l|}{0} & \multicolumn{1}{l|}{1} & \multicolumn{1}{l|}{0} & \multicolumn{1}{l|}{2} & 0 \\ \hline
		\end{tabular}
	}
\end{table*}

Since the SDP program imposes significant limitations, we have turned to a specific quantum state described in the literature that offers an analytical solution, thereby greatly enhancing the efficiency of multi-qubit state generation. This state is known as the n-qubit GHZ-diagonal state. When $n=3$, its form is:

\begin{equation}
	\varrho=\left(\begin{array}{cccccccc}
		\lambda_0 & & & & & & & \mu_0 \\
		& \lambda_1 & & & & & \mu_1 & \\
		& & \lambda_2 & & & \mu_2 & & \\
		& & & \lambda_3 & \mu_3 & & & \\
		& & & \mu_3 & \lambda_3 & & & \\
		& & \mu_2 & & & \lambda_2 & & \\
		& \mu_1 & & & & & \lambda_1 & \\
		\mu_0 & & & & & & & \lambda_0,
	\end{array}\right),
\end{equation}
with $\lambda_i, \mu_i \in \mathbb{R}$.

In Ref. \cite{ghz_d}, the authors propose that for all GHZ-diagonal n-qubit states $\varrho$
\begin{equation}\label{eq.10}
	N_g(\varrho)=\max _i\left\{0,\left|\mu_i\right|-w_i\right\}=\max _i\left\{0, F_i-\frac{1}{2}\right\},
\end{equation}
where $w_i=\sum_{k \neq i} \lambda_k$ and $F_i=\left\langle\psi_i|\varrho| \psi_i\right\rangle$ denotes the fidelity with the GHZ-basis state $\psi_i$. $N_g(\varrho)=\max _i\left\{0, |\mu_i|-w_i\right\}$ holds also true for the slightly more general case with complex $\mu_i$ on the anti-diagonal.

We randomly generate GHZ-diagonal states ranging from 4 to 20 qubits and determine their entanglement status using Eq. (\ref{eq.10}). To balance the samples, we generate 500 quantum states with $-1$ labels and 500 quantum states with $+1$ labels for each qubit. However, when generating high-qubit samples using this method, it becomes easy to generate them but difficult to store them. We only extract the data from the diagonal of each quantum state. Storing 1000 states for 20-qubit quantum states requires 16 GB of storage space. This necessitates software capable of handling large matrices.

For CNN training, we use MATLAB software. When training quantum states above 16 qubits, we utilize the Beijing Super Cloud Computing Center platform. Additionally, as the number of qubits increases, training becomes progressively slower. However, the iteration speed of the algorithm is faster than training GME states, so it is recommended to reduce MaxEpochs to 50.

Before training, we normalize the GHZ-diagonal state. Since it only contains values on the diagonal and the head and tail are equal in size, we can concatenate the data on half of the diagonal into a one-dimensional array. In Fig. \ref{FIG.1}, we use a 5-qubit GHZ-diagonal state, which can be reshaped into a 1$\times$32 matrix and then trained using CNN. Despite this, for a 20-qubit GHZ-diagonal state, the matrix size of 1000 qubit states reaches 1000$\times$1048576.

In Fig. \ref{FIG.3}, we conduct five experiments and obtain the classification results for 4 to 20-qubit quantum states using CNN and CNN-SE, respectively. We observe that both CNN and CNN-SE achieve an average accuracy of over 97$\%$, with CNN-SE exceeding 98$\%$. Specifically, CNN-SE achieves the lowest classification accuracy of 98.07$\%$ for 5-qubits, peaking at 99.80$\%$ for 11-qubits, and maintaining 99.40$\%$ for 20-qubits. Notably, the accuracy remains consistently high for high-dimensional qubits. There seems to be no clear correlation between the number of qubits and the classification accuracy, indicating that the detection of entangled states for GHZ-diagonal states heavily relies on the diagonal values. Thus, the accuracy mainly depends on the CNN network settings and sample quality. This is a favorable situation because the size of the original data hardly affects the classification accuracy of machine learning, which is beneficial for detecting high-dimensional states.

We also examine false positives (non-entangled states predicted as entangled) and false negatives (entangled states predicted as non-entangled). In Table \ref{tab.2}, we tabulate the false positives and false negatives from 9-qubit to 20-qubit states, with 300 states predicted each time. Besides some instances with 0 values, there are numerous single items with 0 values, indicating the partial training perfection of CNN and CNN-SE, as they have no prediction errors. This suggests that the trained model is more polarized, especially in high-dimensional prediction scenarios. After analyzing the results of each qubit, we found that the prediction of entangled states as non-entangled states is more than twice the number of predictions of non-entangled states as entangled states, which is beneficial for the model.

Through these experiments, we demonstrate that the trained model exhibits extremity and is less prone to predicting non-entangled states as entangled states. This situation presents a favorable outcome. Based on the training characteristics of CNN and CNN-SE coupled with their fast training iteration speed, it is feasible to obtain a quantum state classification model that accurately predicts entangled states from non-entangled ones through multiple trainings. This would facilitate our model in classifying more accurately, ensuring the precise detection of all entangled states while maintaining a high accuracy in detecting non-entangled states.

In FIG. \ref{FIG.4}, we present the training process of both CNN and CNN-SE architectures across various qubit dimensions, including 6, 10, 14, and 18 qubits. The depicted loss functions reveal distinct characteristics between the two models and can intuitively reflect the performance of the model under the current parameters. Notably, the CNN model exhibits a swifter convergence rate compared to CNN-SE. However, this accelerated convergence of CNN is accompanied by a susceptibility to overfitting, particularly evident in higher-dimensional scenarios. In contrast, CNN-SE demonstrates a consistent downward trend in fluctuations throughout the training iterations without mutations. This illustrates that the SE attention mechanism we adopt enhances the generalization ability of the model, thus mitigating the harmful effects of overfitting.

\section{DISCUSSION AND CONCLUSION}\label{sec:leve4}

In practical scenarios, quantum systems are invariably subjected to noise. To assess the model's applicability, we investigated the detection of GHZ-diagonal entangled states under white noise influence using machine learning techniques. This study aims to evaluate the robustness and effectiveness of the model in real-world conditions. A GHZ-diagonal quantum state with white noise is
\begin{equation}
	\rho=p\varrho_n+\frac{1-p}{2^n} \mathcal{I},
\end{equation}
where p ranges from 0 to 1, $\varrho_n$ represents the GHZ-diagonal state of n qubits, and $\mathcal{I}$ is the $2^n \times 2^n$ identity matrix.

We randomly generated GHZ-diagonal quantum states ranging from 8 to 12 qubits under the influence of white noise and assessed their entanglement status using Eq. (\ref{eq.10}). To maintain a balanced dataset, we produced 500 entangled states and 500 non-entangled states. Among them, 350 samples of each type were designated for the training set, while the remaining samples were reserved for the test set.

\begin{figure}[t]
	\centering
	% Requires \usepackage{graphicx}
	\includegraphics[width=3.5in]{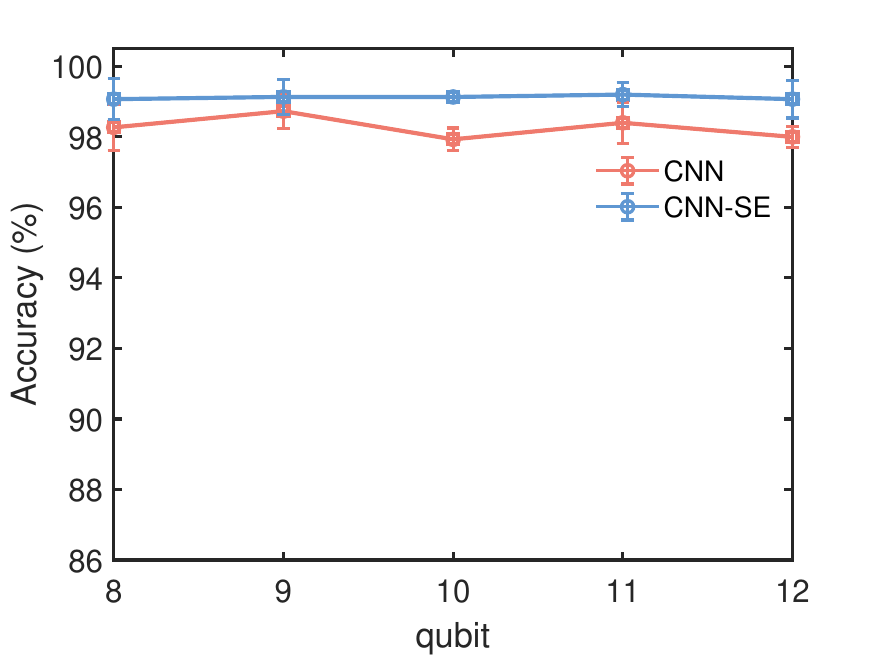}\\
	\caption{Classification accuracy and error bars for GHZ-diagonal states under white noise with different numbers of qubits obtained through CNN and CNN-SE.}\label{FIG.5}
\end{figure}

In Fig. \ref{FIG.5}, we conduct five experiments and obtain the classification results for 8 to 12-qubit quantum states using CNN and CNN-SE, respectively. The experimental results demonstrate that the detection of GHZ-diagonal states under the influence of white noise via CNN and CNN-SE is similar to the unaffected detection results, and both maintain high accuracy. Despite the presence of white noise, our model consistently maintains high accuracy, with CNN-SE generally outperforming CNN in terms of accuracy. These findings underscore the practical effectiveness of our model in handling quantum systems affected by noise.

Extreme models are less apparent in low dimensions, and we aim to enhance polarization while maintaining classification accuracy. Adjusting the threshold \cite{threshold1,threshold2,threshold3} can make the model more conservative or aggressive in prediction, thereby reducing the likelihood of predicting a non-entangled state as an entangled one.

Moreover, improving model accuracy is crucial. The adversarial sample training method \cite{improve1,improve2,improve3} can be employed for this purpose. This approach involves augmenting the training set with adversarial samples, enhancing the dataset and enabling the model to learn from adversarial samples during training. Consequently, the model better adapts to the classification boundary and improves its robustness.

In summary, we employed two algorithms, CNN and CNN-SE, for training and obtained the classification accuracy of the GME state from 4 to 6-qubit and the GHZ-diagonal state from 4 to 20-qubit, respectively. Through comparison, we observed higher classification accuracy with CNN-SE. Furthermore, we discussed the cases of false positives and false negatives. In detecting high-dimensional states, single occurrences were more prevalent. Therefore, through multiple trainings, we can develop a superior model.

\section*{Acknowledgment}
This work is supported by the National Natural Science Foundation of China (Grant No. 11734015, 22103043), and K.C. Wong Magna Fund in Ningbo University.

\bibliographystyle{amsalpha}

\begin{thebibliography}{99}
\bibitem{rev1} M. Horodecki, Entanglement measures, Quant. Inform. Comput. \textbf{1}, 3-26 (2001).

\bibitem{rev2} D. Bru{\ss}, Characterizing entanglement, J. Math. Phys. \textbf{43}, 4237-4251 (2002).

\bibitem{rev3} O. G\"{u}hne, and G. T\'{o}th, Entanglement detection, Phys. Rep. \textbf{474}, 1 (2009).

\bibitem{rev4} M. B. Plenio, and S. Virmani, An introduction to entanglement measures, Quantum Inf. Comput. \textbf{7}, 1-51 (2007).

\bibitem{rev5} R. Horodecki, P. Horodecki, M. Horodecki, and K. Horodecki, Quantum entanglement,  Rev. Mod. Phys. \textbf{81}, 865-942 (2009).

\bibitem{rev6} N. Friis, G. Vitagliano, M. Malik, and M. Huber, Entanglement certification from theory to experiment, Nat. Rev. Phys. \textbf{1}, 72 (2019).

\bibitem{teleportation1} G. Rigolin, Quantum teleportation of an arbitrary two-qubit state and its relation to multipartite entanglement, Phys. Rev. A \textbf{71}, 032303 (2005).

\bibitem{teleportation2} J. Lee, H. Min, and S. D. Oh, Multipartite entanglement for entanglement teleportation, Phys. Rev. A \textbf{66}, 052318 (2002).

\bibitem{teleportation3} P. X. Chen, S. Y. Zhu, and G. C. Guo, General form of genuine multipartite entanglement quantum channels for teleportation, Phys. Rev. A \textbf{74}, 032324 (2006).

\bibitem{coding1} D. Bru{\ss}, M. Lewenstein, A. Sen, U. Sen. G. M. D'ariano, and C. Macchiavello, Dense coding with multipartite quantum states, Int. J. Quantum Inf. \textbf{4}, 415-428 (2006).

\bibitem{coding2} Y. Yeo, and W. K. Chua, Teleportation and dense coding with genuine multipartite entanglement, Phys. Rev. Lett. \textbf{96}, 060502 (2006).

\bibitem{key1} M. Epping, H. Kampermann, and D. Bru{\ss}, Multi-partite entanglement can speed up quantum key distribution in networks, New J. Phys. \textbf{19}, 093012 (2017).

\bibitem{key2} L. Bugalho, B. C. Coutinho, F. A. Monteiro, and Y. Omar, Distributing multipartite entanglement over noisy quantum networks, quantum \textbf{7}, 920 (2023).

\bibitem{Bell1} L. Masanes, Asymptotic violation of Bell inequalities and distillability, Phys. Rev. Lett. \textbf{97}, 050503 (2006).

\bibitem{Bell2} N. Brunner, J. Sharam, and T. Vertesi, Testing the structure of multipartite entanglement with bell inequalities, Phys. Rev. Lett. \textbf{108}, 110501 (2012).

\bibitem{Bell3} X. Wang, C. Zhang, Q. Chen, S. Yu, H. Yuan, and C. H. Oh, Hierarchy of multipartite nonlocality in the nonsignaling scenario, Phys. Rev. A \textbf{94}, 022110 (2016).

\bibitem{cluster} R. Raussendorf and H. J. Briegel, A one-way quantum computer, Phys. Rev. Lett. \textbf{86}, 5188 (2001).

\bibitem{phase1} A. Anfossi, P. Giorda, A. Montorsi, and F. Traversa, Two-point versus multipartite entanglement in quantum phase transitions, Phys. Rev. Lett. \textbf{95}, 056402 (2005).

\bibitem{phase2} T. R. de Oliveira, G. Rigolin, M. C. de Oliveira, and E. Miranda, Multipartite entanglement signature of quantum phase transitions, Phys. Rev. Lett. \textbf{97}, 170401 (2006).

\bibitem{Dicke1} R. H. Dicke, Coherence in spontaneous radiation processes, Phys. Rev. \textbf{93}, 99 (1954).

\bibitem{Dicke2} J. K. Stockton, J. M. Geremia, A. C. Doherty, and H. Mabuchi, Characterizing the entanglement of symmetric many-particle spin-1/2  systems, Phys. Rev. A \textbf{67}, 022112 (2003).

\bibitem{detect1} M. Erhard, M. Krenn, and A. Zeilinger, Advances in high-dimensional quantum entanglement, Nat. Rev. Phys. \textbf{2}, 365-381 (2020).

\bibitem{detect2} O. G{\"u}hne, P. Hyllus, D. Bru{\ss}, A. Ekert, M. Lewenstein, C. Macchiavello, and A. Sanpera, Detection of entanglement with few local measurements, Phys. Rev. A \textbf{66}, 062305 (2002).

\bibitem{detect3} M. Weilenmann, B. Dive, D. Trillo, E. A. Aguilar, and M. Navascu{\'e}s, Entanglement detection beyond measuring fidelities, Phys. Rev. Lett. \textbf{124}, 200502 (2020).

\bibitem{detect4} S. Morelli, H. Yamasaki, M. Huber, and A. Tavakoli, Entanglement detection with imprecise measurements, Phys. Rev. Lett. \textbf{128}, 250501 (2022).

\bibitem{detect5} J. B. Altepeter, E. R. Jeffrey, P. G. Kwiat, S. Tanzilli, N. Gisin, and A. Acín, Experimental methods for detecting entanglement, Phys. Rev. Lett. \textbf{95}, 033601 (2005).

\bibitem{detect6} G. T{\'o}th and O. G{\"u}hne, Detecting genuine multipartite entanglement with two local measurements, Phys. Rev. Lett. \textbf{94}, 060501 (2005).

\bibitem{network1} A. Krizhevsky, I. Sutskever, and G. E. Hinton, ImageNet classification with deep convolutional neural networks, Commun Acm \textbf{60}, 84-90 (2017).

\bibitem{network2} K. Simonyan, and A. Zisserman, Very deep convolutional networks for large-scale image recognition, arXiv:1409.1556.

\bibitem{network3} K. He, X. Zhang, S. Ren, and J. Sun, Deep residual learning for image recognition, Proc. IEEE Conf. Comput. Vis. Pattern Recog., 770-778 (2016).

\bibitem{network4} L. Zhang, Z. Chen, and S. M. Fei, Entanglement verification with deep semisupervised machine learning, Phys. Rev. A \textbf{108}, 022427 (2023).

\bibitem{svm} C. Cortes, and V. Vapnik, Support vector machine, Mach. Learn. \textbf{20}, 273-297 (1995).

\bibitem{s4vm} Y. Li, and Z. Zhou, Improving semi-supervised support vector machines through unlabeled instances selection, Proc AAAI Conf Artif Intell. \textbf{25}, 386-391 (2011).

\bibitem{entanglement1} Y. J. Luo, J. M. Liu, and C. Zhang, Detecting genuine multipartite entanglement via machine learning, Phys. Rev. A \textbf{108}, 052424 (2023).

\bibitem{entanglement2} S. Lu, S. Huang, K. Li, J. Li, J. Chen, D. Lu, Z. Ji, Y. Shen, D. Zhou, B. Zeng, Separability-entanglement classifier via machine learning, Phys. Rev. A \textbf{98}, 012315 (2018).

\bibitem{entanglement3} Y. Levine, O. Sharir, N. Cohen, and A. Shashua, Quantum entanglement in deep learning architectures, Phys. Rev. Lett. \textbf{122}, 065301 (2019).

\bibitem{entanglement4} Y. Chen, Y. Pan, G. Zhang, and S. Cheng, Detecting quantum entanglement with unsupervised learning, Quantum Sci. Technol. \textbf{7}, 015005 (2021).

\bibitem{entanglement5} C. Chen, C. Ren, H. Lin and H. Lu, Entanglement structure detection via machine learning, Quantum Sci. Technol. \textbf{6}, 035017 (2021).

\bibitem{steer1} L. Zhang, Z. Chen, and S. M. Fei, Einstein-Podolsky-Rosen steering based on semisupervised machine learning, Phys. Rev. A \textbf{104}, 052427 (2021).

\bibitem{steer2} C. Ren and C. Chen, Steerability detection of an arbitrary two-qubit state via machine learning, Phys. Rev. A \textbf{100}, 022314 (2019).

\bibitem{steer3} G. Z. Pan, M. Yang, J. Zhou, J. Zhou, M. Kong, and G. Zhang, Optimization of tripartite quantum steering inequalities via machine learning, Quantum Inf. Process \textbf{22}, 162 (2023).

\bibitem{steer4} H. M. Wang, H. Y. Ku, J. Y. Lin, and H. B. Chen, Deep learning the hierarchy of steering measurement settings of qubit-pair states, Commun. Phys. \textbf{7}, 72 (2024).

\bibitem{sdp} M. X. Goemans, and D. P. Williamson, Improved approximation algorithms for maximum cut and satisfiability problems using semidefinite programming, J. ACM \textbf{42}, 1115-1145 (1995).

\bibitem{machine1} F. Galton, Regression towards mediocrity in hereditary stature, J. Anthropol. Inst. Great Brit. Ireland. \textbf{15}, 246-263 (1886).

\bibitem{machine2} K. Pearson, LIII. On lines and planes of closest fit to systems of points in space, Philos. Mag. J. Sci. \textbf{2}, 559-572 (1901).

\bibitem{machine3} M. Minsky, and S. A. Papert, Perceptrons, reissue of the 1988 expanded edition with a new foreword by L{\'e}on Bottou: an introduction to computational geometry, MIT press, (2017).
	
\bibitem{machine4} K. Fukushima, Neocognitron: A self-organizing neural network model for a mechanism of pattern recognition unaffected by shift in position, Biol. Cybern. \textbf{36}, 193-202 (1980).
	
\bibitem{machine5} J. L. McClelland, D. E. Rumelhart, and PDP Research Group, Parallel distributed processing, MIT press, (1987).
	
\bibitem{machine6} D. E. Rumelhart, G. E. Hinton, and R. J. Williams, Learning representations by back-propagating errors, nature \textbf{323}, 533-536 (1986).
	
\bibitem{machine7} T. Brown, B. Mann, N. Ryder, et al, Language models are few-shot learners, Adv. Neural Inf. Process. Syst. \textbf{33}, 1877-1901 (2020).
	
\bibitem{machine8} W. Fedus, B. Zoph, and N. Shazeer, Switch transformers: Scaling to trillion parameter models with simple and efficient sparsity, J. Mach. Learn. Res. \textbf{23}, 1-39 (2022).

\bibitem{machine9} D. Narayanan, M. Shoeybi, J. Casper, et al, Efficient large-scale language model training on gpu clusters using megatron-lm, Proc. Int. Conf. High Perform. Comput. Netw. Storage Anal., 1-15 (2021).

\bibitem{nonlocality1} P. Broecker, F. Assaad, and S. Trebst, Quantum phase recognition via unsupervised machine learning, arXiv:1707.00663.

\bibitem{nonlocality2} A. Canabarro, S. Brito, and R. Chaves, Machine learning nonlocal correlations, Phys. Rev. Lett. \textbf{122}, 200401 (2019).

\bibitem{spin1} Y. Zhang, R. G. Melko, and E. A. Kim, Machine learning Z2 quantum spin liquids with quasiparticle statistics, Phys. Rev. B \textbf{96}, 245119 (2017).

\bibitem{spin2} Y. H. Teoh, M. Drygala, R. G. Melko, and R. Islam, Machine learning design of a trapped-ion quantum spin simulator, Quantum Sci. Technol. \textbf{5}, 024001 (2020).

\bibitem{spin3} W. J. Rao, Machine learning the many-body localization transition in random spin systems, J. Phys-condens Mat. \textbf{30}, 395902 (2018).

\bibitem{cnn} Y. LeCun, L. Bottou, Y. Bengio, and P. Haffner, Gradient-based learning applied to document recognition, Proc. IEEE \textbf{86}, 2278-2324 (1998).

\bibitem{se} J. Hu, L. Shen, and G. Sun, Squeeze-and-excitation networks, IEEE Conf. Comput. Vis. Pattern Recognit., 7132-7141(2018).
	
\bibitem{ghz_d} M. Hofmann, T.Moroder, and O. G{\"u}hne, Analytical characterization of the genuine multiparticle negativity, J. Phys. A: Math. Theor. \textbf{47}, 155301 (2014).

\bibitem{gmn} B. Jungnitsch, T. Moroder, and O. G{\"u}hne, Taming multiparticle entanglement, Phys. Rev. Lett. \textbf{106}, 190502 (2011).

\bibitem{adam1} D. P. Kingma, and J. Ba, Adam: A method for stochastic optimization, arXiv:1412.6980.

\bibitem{adam2} N. Qian, On the momentum term in gradient descent learning algorithms, Neural Netw. \textbf{12}, 145-151 (1999).

\bibitem{adam3} X. Glorot, and Y. Bengio, Understanding the difficulty of training deep feedforward neural networks, Proc. 13th Int. Conf. Artif. Intell. Stat. \textbf{9}, 249-256 (2010).

\bibitem{threshold1} B. Zhou, A. Khosla, A. Lapedriza, A. Oliva, and A. Torralba, Learning deep features for discriminative localization, Proc. IEEE Conf. Comput. Vis. Pattern Recog., 2921-2929 (2016).

\bibitem{threshold2} T. Y. Lin,  P. Goyal, R. Girshick, K. He, and P. Doll{\'a}r, Focal loss for dense object detection, IEEE Int. Conf. Comput. Vis., 2980-2988 (2017).

\bibitem{threshold3} T. DeVries, and G. W. Taylor, Learning confidence for out-of-distribution detection in neural networks, arXiv:1802.04865.

\bibitem{improve1} I. J. Goodfellow, J. Shlens, and C. Szegedy, Explaining and harnessing adversarial examples, arXiv:1412.6572.
	
\bibitem{improve2} T. Miyato, A. M. Dai, and I. Goodfellow, Adversarial training methods for semi-supervised text classification, arXiv:1605.07725.
	
\bibitem{improve3} A. Shafahi, M. Najibi, M. A. Ghiasi, et al, Adversarial training for free!, Adv. Neural Inf. Process. Syst. \textbf{32}, (2019).
	
	
	
	
	
\end{thebibliography}

\end{document}